\documentclass{llncs}
  
\usepackage{graphicx}
\usepackage{subfigure}

\usepackage{amssymb}
\usepackage{amsmath}

\usepackage{url}
\usepackage[dvipsnames]{xcolor}

\makeatletter
\newif\if@restonecol
\makeatother

\usepackage{multirow}
\usepackage{mathrsfs}

\usepackage{url}
\usepackage{calc}

\usepackage{array} 
\usepackage{pdflscape}

\allowdisplaybreaks

\begin{document}
\title{Organic Computing in the Spotlight}
\author{Sven Tomforde, Bernhard Sick, and Christian M\"uller-Schloer}
\institute{Intelligent Embedded Systems Lab, University of Kassel, Germany\\
	Institute for Systems Engineering, Leibniz University of Hannover\\
\email{$\{$stomforde$|$bsick$\}$@uni-kassel.de;~cms@sra.uni-hanover.de}}

\maketitle

\begin{abstract}
	Organic Computing is an initiative in the field of systems engineering that proposed to make use of concepts such as self-adaptation and self-organisation to increase the robustness of technical systems. Based on the observation that traditional design and operation concepts reach their limits, transferring more autonomy to the systems themselves should result in a reduction of complexity for users, administrators, and developers. However, there seems to be a need for an updated definition of the term ``Organic Computing'', of desired properties of technical, organic systems, and the objectives of the Organic Computing initiative. With this article, we will address these points.
\end{abstract}


\section{Introduction}

Within the last decade, information and communication technology (ICT) has witnessed a dramatic change: from isolated manually controlled systems to highly interconnected and increasingly autonomous systems \cite{THS14,TRBW16}. This change is a reaction to the observation that traditional engineering and control concepts have reached their limits: In a complex world of perpetual change and uncertainty, static design-time decision processes cannot appropriately anticipate runtime conditions and, consequently, cannot foresee suitable reactions. We still have to cope with heavily increasing system complexity.

In order to provide mechanisms to cope with these complexity challenges, initiatives such as Organic Computing \cite{MSMW04} have been proposed. The overall goal of Organic Computing (OC) is nothing less than a paradigm shift in systems engineering. Such a new paradigm should enable future ICT systems to carry out certain tasks on their own, they should have properties such as self-organisation, self-healing, self-configuration, self-protection, self-optimisation, and they should be able to adapt reasonably to changing requirements of their operating environment \cite{TP+11}. In particular, they should adjust to human needs in a trustworthy way and should allow for explicit interference in case of undesired effects of self-organisation, leading to the---somewhat contradictory---requirement of controlled self-organisation. Introducing self-* properties entails a dramatic change in the responsibilities: It means to move traditional design-time decisions to runtime and from the system engineer to the systems themselves. This transfer of control has impact on all phases of the design process: from initial requirement specification to design and development, and finally to the maintenance of systems.

It is a misconception that the goal of building organic systems is primarily the construction of self-adaptive and/or self-organising systems. Self-organisation and self-adaptation are just means to make technical systems resistant against external or internal disturbances. It is also a misconception to assume that organic systems (or self-adaptive and self-organising systems in general) generally achieve a higher performance (e.g., higher speed or better quality of decisions) than conventional systems. Organic Computing systems are not per se faster than conventional systems but they return faster to a certain corridor of an acceptable performance in the presence of disturbances. The ultimate goal of organic systems is to become more resilient against disturbances and attacks from outside. We call this property ``robustness'' \cite{SM+10}.

Although several contributions from the state of the art discussed the motivation, the ideas, and key technology of Organic Computing, there is no clear definition of the term. Furthermore, we often face a misconception in the sense that the goal of Organic Computing is to develop self-organising systems. But mechanisms such as self-organisation and adaptivity are rather the means and not the end. The ultimate goal of OC systems is to equip ICT systems with higher degrees of robustness in the presence of external and internal disturbances. With this article, we propose a renewed definition of the term ``Organic Computing'' (Section~\ref{sec:term}) and explain the desired properties an organic system should have (Section~\ref{sec:properties}). Furthermore, we give a brief summary of the history of Organic Computing (Section~\ref{sec:history}) and give an outlook to urgent research challenges (Section~\ref{sec:outlook}). 

\section{Term Definition}
\label{sec:term}

Organic Computing is based on the insight that we are increasingly surrounded by large collections of autonomous systems, which are equipped with sensors and actuators, aware of their environment, communicate freely, and organise themselves in order to perform the actions and services that seem to be required. Consequently, we define the term ``Organic Computing'' (OC) as follows:

An ``Organic Computing System'' (or just an organic system) is a technical system, which is equipped with sensors (to perceive its environment) and actuators (to manipulate it). It adapts autonomously and dynamically to the current conditions of the perceived environment. This adaptation process has impact on the system's utility, which is continuously improved by the organic system itself. To allow for such an adaptive behaviour, it employs so-called self-* mechanisms. 

Typically, an organic system consists of a potentially large set of such autonomous organic subsystems. Conceptually, each OC system contains two complementary parts: one is responsible for the productive operation of the system (i.e, it fulfils the technical purpose) and one for the adaptation aspects (i.e., realising the organic capabilities). In order to react to previously unknown and unanticipated conditions with appropriate behaviour, an organic system is typically based on (machine) learning techniques.

At this point, some clarification of the term ``Organic Computing'' seems to be useful. The adjective ``organic'' has several meanings. The Cambridge Dictionary\footnote{See \url{http://dictionary.cambridge.org/} (last access 2016/13/12)} lists three of them: ``of or concerning organs of the body'', ``of, found in, or produced by, living things'', and ``(of food) grown without the use of artificial fertilisers''. From the previous discussion, the second meaning is the one relevant for Organic Computing: It aims at augmenting technical systems with properties that are similar to those found in ``living things''. Organic Computing is about engineering technical systems, but it does not mean to build them out of organic material or living organisms! ``Organic'' has also a connotation reminding of ``organisation'': Organic Computing deals with ways how to organise systems consisting of systems with increasing autonomy.

It is the objective of Organic Computing to equip technical systems with ``life-like'' properties. This includes, in the first place, robustness, flexibility, and resilience against disturbances and attacks. However, in the second place, it also includes avoidance of single-points-of-failures, a decrease of control and management complexity, and more customised behaviour.

\section{Self-* Properties of Organic Systems}
\label{sec:properties}

Occasionally, Organic Computing is equated with self-organisation. However, it is a misconception that the term ``self-organisation'' comprises all aspects of systems with a high degree of autonomy. ``Self-organisation'' is just one mechanism which eventually leads to a system with self-* properties.

The attribute ``self'' refers to a system's autonomy, meaning there is no or only limited (or goal-oriented) external control guiding the system's behaviour. Autonomy can be formalised using the concept as discussed in \cite{SM+10}: There, a \textit{degree of autonomy} is introduced that allows to define a range of control options: from total autonomy (i.e., no external control) to no autonomy (all control is exerted from the outside)---and intermediate degrees of autonomy where internal (self-)control and external control are combined. For the internal (self) part of control, an observer/controller unit \cite{TP+11} within the OC system is responsible. We want to emphasise that neither full autonomy nor zero autonomy are desirable: While we want to build systems that can take local decisions themselves we do not want fully autonomous systems that cannot controlled by the human user any longer! 

Although researchers in the field agree on the need of self-* properties, the definitions of the individual properties largely differ---if a clear definition is given at all. In the following section we aim at providing a taxonomy of the most important self-* properties in the context of Organic Computing.

In literature (see e.g. \cite{MW+07}), the terms self-management and structural adaptivity are used. ``Self-management'' is especially popular in the Autonomic Computing community \cite{KC03}. For clarification, we suggest the following definition, which aims at being compatible with the general usage of the terms:
\begin{enumerate}
	\item Systems can be modified at runtime in terms of (i) structure and/or (ii) behaviour. Structure defines the components and the connections among components or subsystems, while behaviour describes the sequence of events (and actions). ``Behaviour'' can also be seen as the process running on and between the components. However, all of the following properties have impact on (at least) the system's behaviour.
	\item \textbf{Self-configuration} (often also called \textit{self-adaptation}) is used to modify the parameters of the system. Organic systems configure themselves (by parametrisation resulting in a behaviour) in accordance with higher-level user goals.
	\item \textbf{Self-organisation} relates to a change of the structure of a system (i.e., of components or subsystems and their links). An organic system validates the status of relations towards other subsystems continuously at runtime based on the currently active user goal. 
	\item \textbf{Self-integration} is related to both aspects (i.e., self-configuration and self-organisation) in the context of combination and interplay of several systems. The organic system decides autonomously about its role within the overall system and adapts its behaviour as well as the relations to others (i.e, its sphere of influence regarding the system's structure) to function correctly as an element of the overall system.\footnote{Self-integration becomes increasingly important in the context of ``Interwoven Systems'' (see, e.g., \cite{THS14,TRBW16,HBLMST15}), i.e., open, large-scale, and distributed systems that evolve during runtime and dynamically combine heterogeneous subsystems that belong to varying authorities with a possibly different application field.}
	\item \textbf{Self-management} comprises self-configuration, self-organisation, and possibly further self-* mechanisms.
\end{enumerate}

\noindent Further self-* mechanisms that are particularly important for organic systems are:

\begin{itemize}
	\item \textbf{Self-healing}: Organic systems are able to detect, diagnose, and repair failures. This includes a localisation of these failures (in both, software and hardware), possibly using external sources through cooperation with other systems and their sensory equipment.
	\item \textbf{Self-protecting}: Organic systems have to protect themselves in terms of defending the system (i.e., the system itself as well as the overall system it is part of) against attacks. Besides attacks from outside, this also includes large-scale cascading failures that go beyond the scope of the self-healing mechanism (i.e., affect the overall system rather than the organic system under consideration). Protection further requires proactive mechanisms that try to identify behaviour and conditions that may result in failures---and work against them before they become pathological.
	\item \textbf{Self-stabilising}: Organic systems face the challenge to come up with a stable behaviour: Either due to continuous adaptation processes or due to external effects caused by the influences of other subsystems. An organic system is able to suppress such unstable behaviour over time. 
	\item \textbf{Self-improving} (often also called \textit{self-optimisation}\footnote{We favour the term ``improvement'' over ``optimisation'', since optimisation indicates to search for the optimum and converge to this solution. However, organic systems have to act under real-world conditions, i.e., in real-time and with limited knowledge. This means they often have to come up with ``good-enough'' solutions rather than spending high effort in trying to find \textit{the optimal} one (which might be hardly achievable in most systems).}) means that the organic system continuously analyses its decisions with the goal to find better solutions. In particular, this means that all previous mechanisms (especially self-configuration and self-organisation) become subject to an improvement by learning at runtime.
	\item \textbf{Self-explaining:} The implementation of OC ideas leads to systems with higher degrees of autonomy. But the (human) user must always keep the ultimate control over the system although she or he is not involved in low-level detail decisions. Since the OC system has certain degrees of freedom regarding its decisions it must be able to reason about them upon request, and must ask for help. This is often true in situations where a multi-objective optimisation is required.
\end{itemize}

An alternative line of classification is to consider *-awareness. In this context, ``*'' is used as placeholder for the particular scope, i.e., ``self'', ``environment'', or ``context'' to name just the most prominent aspects.

%
%
%
%
%

\section{A Brief History of Organic Computing}
\label{sec:history}

The term ``Organic Computing'' has been coined in the context of a workshop on future topics in computer engineering, which has been organised in November 2002 by the special interest group ``Computer Architecture'' (Fachausschuss ARCS -- Architektur von Computer-Systemen) within the German Computer Science Society (Gesellschaft f\"ur Informatik, GI). The goal of this workshop was to predict some important technical developments in computer science for the next 5 to 10 years and to derive the corresponding challenges for industry and academia. The results of the workshop have been summarised under the headline ``Organic Computing'' in a position paper of GI and ITG (Informationstechni\-sche Gesellschaft -- information-technical society), which has been authored by 20 representatives from academia and industry working in the field of technical computer science \cite{GI}.

Independently of and in parallel to this origin of the term ``Organic Computing'', an initiative mainly influenced by Christoph von der Malsburg came up with the same term \cite{Mal99}. He organised a workshop already in 2001 that discussed challenges in this context. In contrast to the definition postulated by this book (and in contrast to the research efforts in the field of Organic Computing subsequently to the initial workshops), the term there has been used to combine neuro sciences, molecular biology, and software engineering. The focus of interest was on developing novel computing systems that make use of biological principles and method. 

In 2002, Forrester Research presented a study that postulated to use ``Organic IT'' as novel strategy in information system structures \cite{GR+02}. This vision has subsequently been supported by a number of large companies, including HP, IBM, and Microsoft. As one of the consequences, novel adaptive system architectures have been developed in this environment. 

Closely related and almost in parallel to the Organic Computing initiative, the Autonomic Computing initiative mainly driven by IBM emerged \cite{KC03}. Inspired by the functionality of the autonomic nervous system in human beings, a management solution for large-scale computing and data centres was proposed that autonomously and self-sufficiently reacts on changes in the system's environment to maintain the desired system utility. Such a process requires a large degree of self-organisation. For instance, disturbances are countered in a self-healing way to recover the system from undesired states. This feature of ``self-healing'' is just one of several self-* properties that are used to characterised autonomic systems, the most prominent ones being self-configuration, self-optimisation, and self-protection. These properties are considered to be mandatory to be able to master the growing complexity in technical development.

In close timely vicinity, further ideas to equip future computing systems with self-* properties emerged. Besides the German Research Foundation (DFG) Priority Programme on Organic Computing (see below), this entailed several closely related funding programmes. For instance, investigations on complex systems and concepts for developing such systems have been financed by the European Union in the context of the ``Future and Emergent Technologies'' programme or the Australian CSIRO Centre for Complex Systems.

\begin{figure}[ht!]
\centering
\includegraphics[width=0.99\linewidth]{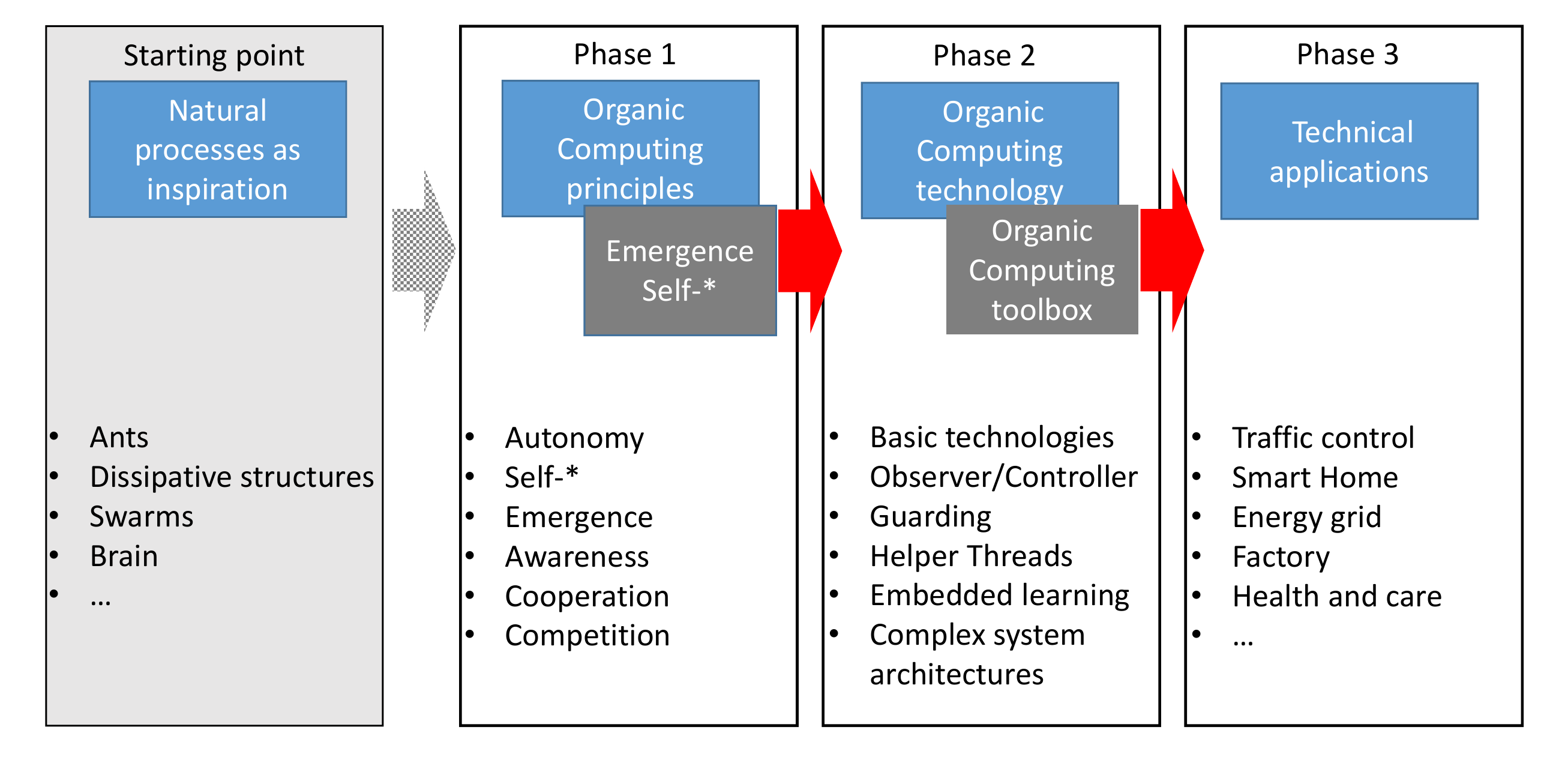}
\caption{Schematic organisation of the priority programme 1183.}
\label{fig:SPPphases}
\end{figure}

The main driving force behind the Organic Computing initiative was the Priority Programme 1183, financed by the DFG from 2005 to 2011. The results of this programme have been summarised in \cite{MSSU11OC}, a intermediate collection of organic technology was published in \cite{Wue08}. The priority programme was composed of three consecutive phases of two years each, Figure~\ref{fig:SPPphases} illustrates the organisation. In the first phase, a deeper analysis and understanding of self-organisation processes has been performed. This has been expected to result in a terminological framework, quantification mechanisms for major aspects of Organic Computing systems, and a basic concept of how such systems are designed and operated. Afterwards, the second phase was intended to shift the focus towards insights on developing Organic Computing systems, including the basic technology to equip such systems with the desired self-* properties. A deliverable that was assumed to be subsequently extended was an Organic Computing toolbox that combined all transferable and generalisable software solutions developed within the priority programme. Finally, the third phase aimed for transferring the derived knowledge into technical applications. However, this organisation into three phases was the initial plan, and the underlying research is still an ongoing process.

After finishing the priority programme, several activities have been started to keep the OC community together and jointly shed even more light into the field of Organic Computing. As one particular example, an OC perspectives workshop series has been performed in 2013 and 2014 in H\"unfeld, Germany, where researchers from Switzerland and Germany came together to discuss the upcoming challenges for the Organic Computing community.

In addition, two workshop series have been established that both provide a yearly basis for exchange and discussion of novel insights in the field: The International Workshop on Self-Optimisation in Autonomic and Organic Computing Systems (SAOS) \cite{ARCS-W14,ARCS-W15,ARCS-W16} is co-located with the International Conference on Architecture of Computing Systems (ARCS) and will see its fifth edition in 2017 in Vienna, Austria; the International Workshop on Self-Improving System Integration (SISSY) is either co-located with the ACM/IEEE International Conference on Autonomic Computing (ICAC) \cite{BellmanTW16} or the IEEE International Conference on Self-Adaptive and Self-Organising Systems (SASO) \cite{BTW14}. The upcoming event in 2017 is the fourth instance and will take place in Columbus, Ohio, USA.

Finally, a series of Organic Computing Doctoral Dissertation Colloquia (OC-DDC) has been established \cite{Tom13,TS14,TS15}, which will take place in its fifth edition in June 2017 in Bochum, Germany.

\section{Outlook: Organic Computing Development Options}
\label{sec:outlook}
Since the conception of the idea in 2002, OC has progressed in different directions. While many theoretical and practical results have been achieved, it became quite clear during this time that there is much more effort needed: OC is a long-term research project. Within the limited space of this article, we can only briefly sketch a few development options:
\begin{itemize}
	\item \textbf{Design-time to runtime:} OC can be characterised by the motto: ``Move design-time decisions into runtime!''. Any autonomous modification at runtime is subject to strict real-time and safety constraints which necessitates a rethinking of traditional design-time activities such as modelling, design space exploration (which becomes configuration space exploration), validation (by simulation), verification, testing etc.
	\item \textbf{Optimisation at runtime:} OC systems can be understood as large distributed multi-element systems where each sub-system tries permanently to improve itself (in the sense of a ``good enough'' solution) and where the collection of sub-systems tries to optimise on the system level as well. Individual and system goals might coincide or diverge. The optimisation problems the OC system faces are usually blackbox problems which cannot be described in a closed form. We need methods that can explore such unknown fitness landscapes, learn their structure and navigate in this landscape within time frames still suitable for real-time reactions.
	\item \textbf{Social OC:} OC systems can be understood as agents interacting with each other in large collectives. These collectives are usually open, i.e., additional members can join at any time. This means that we do not know the internals of the agents. Agents can be trustworthy and co-operative, but they can as well misbehave unintentionally or on purpose. Social mechanisms such as trust, reputation, or fairness have been already successfully employed to help such communities self-organise and improve their overall system performance. There exist more social mechanisms such as, e.g., norms or rule systems such as those of Enduring Institutions \cite{Ost90} that could be transferred to technical systems.
	\item \textbf{Systems of systems:} OC systems will usually not be built from scratch; they will more often be assembled from existing systems. This poses the question of system integration at runtime. We have to clarify the relationship of sub-systems on the same level (peer-to-peer) but also between higher-level and lower-level systems in a hierarchy. If systems are nested within other systems: What is the relationship of the contained subsystems to their superiors, or to other contained sub-systems? The idea of regarding sub-systems as holons (i.e., two-faced system elements with one role of a part and one role of a whole, see \cite{DFMPT16}) linked to other sub-systems via well-defined goal-response relationships is a promising approach with great potential for the task of runtime system integration.
	\item \textbf{Autonomous learning:} An essential property of OC systems is (machine) learning. In contrast to other research domains, learning processes have to be utilised in a highly autonomous fashion. In particular, sophisticated pre-training, parameter studies to determine appropriate configurations of learning mechanisms, and continuous user feedback are not an option. Here, current developments in the field of machine learning that explicitly focus on autonomy are of potential benefit \cite{HT15}.
\end{itemize}

As usual in complex systems and OC, these development options are not independent of each other: Holonic goal-oriented approaches have a strong social component, the agents strive for improvement or even optimisation individually and/or collectively, and they use methods for configuration space exploration, runtime modelling and simulation etc.

\section{Conclusion}
\label{sec:conclusion}

Within this article, we briefly summarised the historical development in the context of the Organic Computing initiative. Organic Computing is an initiative of systems engineering that proposed to make use of concepts such as self-adaptation and self-organisation to increase the robustness of technical systems. Based on the observation that traditional design and operation concepts reach their limits, transferring more autonomy to the systems themselves should result in a reduction of complexity for users, administrator, and developers. Since---at least from our point of view---no precise and commonly accepted notion of the term is available in literature, we derived a specification of the term and the underlying objectives of the initiative, and augmented this with an explanation of the most important self-* properties of OC systems. Finally, future development options of OC have been sketched.

\bibliographystyle{splncs}
\bibliography{bib/OC}

\end{document}